\documentclass[reprint, prb, twocolumn,floatfix]{revtex4}
\usepackage[latin1]{inputenc}
\usepackage{enumerate}
\usepackage{amsfonts}
\usepackage{amssymb}
\usepackage{amsmath,amsbsy}
\usepackage{dcolumn}
\usepackage{bm}
\usepackage{graphicx, graphics}
\usepackage{color}

\DeclareMathOperator*{\ketGS}{\left|\psi_{0}\right\rangle}
\DeclareMathOperator*{\braGS}{\left\langle \psi_{0}\right|}

\begin{document}
		\title{Quantum phase transitions detected by a local probe using time correlations and violations of Leggett-Garg inequalities}
		\author{F. J. Gómez-Ruiz$^1$}
		\email{ fj.gomez34@uniandes.edu.co}
		\author{J. J. Mendoza-Arenas$^{1,2}$}
		\author{F. J. Rodríguez$^1$}
		\author{C. Tejedor$^3$}
        \author{L. Quiroga$^1$}
		\affiliation{$^1$Departamento de Física, Universidad de los Andes, A.A. 4976, Bogotá D. C., Colombia}
		\affiliation{$^2$Clarendon Laboratory, University of Oxford, Parks Road, Oxford OX1 3PU, United Kingdom}
		\affiliation{$^3$Departamento de Física Teórica de la Materia Condensada and Condensed Matter Physics Center (IFIMAC), Universidad Autónoma de Madrid, 28049, Spain}		
		\date{\today}
		\begin{abstract}
In the present paper we introduce a way of identifying quantum phase transitions of many-body systems by means of local time correlations and Leggett-Garg inequalities. This procedure allows to experimentally determine the quantum critical points not only of finite-order transitions but also those of infinite order, as the Kosterlitz-Thouless transition that is not always easy to detect with current methods. By means of simple analytical arguments for a general spin-$1 / 2$ Hamiltonian, and matrix product simulations of one-dimensional $X X Z$ and anisotropic $X Y$ models, we argue that finite-order quantum phase transitions can be determined by singularities of the time correlations or their derivatives at criticality. The same features are exhibited by corresponding Leggett-Garg functions, which noticeably indicate violation of the Leggett-Garg inequalities for early times and all the Hamiltonian parameters considered. In addition, we find that the infinite-order transition of the $X X Z$ model at the isotropic point can be revealed by the maximal violation of the Leggett-Garg inequalities. We thus show that quantum phase transitions can be identified by purely local measurements, and that many-body systems constitute important candidates to observe experimentally the violation of Leggett-Garg inequalities.
			\end{abstract}
		\maketitle

\section{Introduction} \label{intro}

In recent years, quantum phase transitions (QPTs) of many-body systems have been the object of intense research~\cite{sachdev,aeppli}. This is the case not only due to the intrinsic interest that critical phenomena exhibit but also because the understanding and development of new states in condensed matter or atomic systems may have prominent applications in areas such as high-temperature superconductivity~\cite{lee2006rmp} and quantum computation~\cite{nielsen}. The seminal recent advances on quantum simulation schemes~\cite{georgescu2014rmp} in systems such as cold atoms in optical lattices~\cite{bloch2008rmp} and trapped ions~\cite{islam2011nat,schneider2012rpp} constitute fundamental steps in this direction.

Usually finite-order QPTs of a particular system are characterized by discontinuities of its ground state energy or singularities of its derivatives with respect to the parameter that drives the transitions. Besides the determination of order parameters, quantities such as gaps, spatial correlation functions, and structure factors are commonly used to determine the quantum critical points of several models. Remarkably, a few years ago it was realized that entanglement plays a fundamental role in critical phenomena and that different measures of entanglement can be used to determine the location of several types of QPTs~\cite{amico2002nature,gu2003pra,latorre2003prl,wu2004prl,mosseri2004pra,Reslen2005epl,laflorencie2006prl,zanardi2006njp,amico2008rmp,buonsante2007prl,yo2010pra,canovi2014prb,hofmann2014prb}. Furthermore, the relation of Bell inequalities and criticality has been recently explored~\cite{justino2012pra,sun2014pra,sun2014_2pra}.

Since nonlocal measurements are not always accessible, in this paper we propose an alternative form to characterize QPTs by exploiting single-site protocols to obtain bulk properties of many-body systems~\cite{Gessner2011PRL,Gessner2013PRA,Gessner2014epl,Gessner2014Nat}. We argue that local time correlations can indicate the location of critical points for finite-order QPTs, in a similar way to measures of bipartite entanglement such as concurrence and negativity~\cite{wu2004prl}. This is exemplified by numerical simulations, based on tensor network algorithms, of time correlations of one-dimensional (1D) spin-$1 / 2$ lattices described by $X X Z$ and anisotropic $X Y$ Hamiltonians, which correspond to exhaustively-studied models of condensed matter physics. The first- and second-order transitions of these models are determined by nonanalyticities of the time correlations and their first derivative, respectively. We also relate QPTs to a different characterization of quantumness of a system, namely the violation of Leggett-Garg inequalities (LGI)~\cite{lgi_original,emary2014rpp,huelga1996pra,castillo2013pra,bell_leggett_garg_2015,kofler2007prl}, which indicates the absence of macroscopic realism and noninvasive measurability. We show that by maximizing the violation of these inequalities along all possible directions, the infinite-order QPT of the $X X Z$ model can be identified. Given that the models considered in our work describe several condensed-matter systems~\cite{aeppli} and can be implemented in a variety of quantum simulators~\cite{georgescu2014rmp}, our analysis places them as interesting many-body scenarios for the experimental observation of the violation of Leggett-Garg inequalities.

The paper is organized as follows: Section~\ref{sect_qpt} describes the 1D spin models we focus on, whose QPTs are well known and thus allow us to check the adequacy of our proposal. Section~\ref{section_time_correl} discusses how finite-order QPTs can be identified from local time correlations, providing examples for the spin models previously described. Leggett-Garg inequalities and their role for determining QPTs are analyzed in Sec.~\ref{sect_lgi}. Finally, Sec.~\ref{conclu} contains the conclusions drawn from our work.
\section{Quantum Phase Transitions in One-Dimensional Spin-$1 / 2$ models} \label{sect_qpt}
The main goal of our work corresponds to determining whether local unequal-time correlations and Leggett-Garg inequalities can be used to localize and characterize QPTs. To provide an answer to this problem, we analyze the time correlations of systems described by spin-$1 / 2$ time-independent Hamiltonians of the form
\begin{equation} \label{Hami_general}
H=\sum_{\alpha}\sum_{i,j} J_{\alpha}^{i,j}\sigma_i^{\alpha}\sigma_j^{\alpha}+\sum_{\alpha}\sum_{i}B_{\alpha}^{i}\sigma_i^{\alpha}.
\end{equation}
Here $\sigma_i^{\alpha}$ denotes the Pauli operators at site $i$ ($\alpha=x,y,z$), $J_{\alpha}^{i,j}$ is the coupling between spins at sites $i$ and $j$ along direction $\alpha$, $B_{\alpha}^{i}$ is the magnetic field at site $i$ along direction $\alpha$, and $\hbar=1$. No restrictions on the dimensionality of the system or the range of the interactions are in principle required.

While our analytical arguments are based on the Hamiltonian of Eq.~\eqref{Hami_general} and thus are quite general (see Appendix~\ref{proof}), we restrict our numerical studies to two particular testbed Hamiltonians of condensed matter physics, namely the 1D $X X Z$ and anisotropic $X Y$ models with nearest-neighbour interactions. These systems have been extensively studied in the literature, and their ground-state phase diagrams are very well known~\cite{takahashi,giamarchi,Sutherland}. In this section we briefly describe the QPTs featured by these models.

\subsection{Spin-$\frac{1}{2}$ X X Z Model}
We first consider a 1D system in which $1 / 2$  spins are coupled through an anisotropic Heisenberg interaction. This case, known as the $X X Z$ model, corresponds to $J_{x}^{i,j}=J_{y}^{i,j}=J$, $J_{z}^{i,j}=J\Delta$, and $B_{\alpha}^i=0$. Thus it is described by the Hamiltonian
\begin{equation} \label{hami_xxz}
H=J\sum_{i}\left[\sigma_{i}^{x}\sigma_{i+1}^{x} + \sigma_{i}^{y}\sigma_{i+1}^{y} +\Delta \sigma_{i}^{z}\sigma_{i+1}^{z}\right].
\end{equation}		
Here the coupling $J > 0$ represents the exchange interaction between nearest neighbors, and $\Delta$ is the dimensionless anisotropy along the $z$ direction~\footnote{Alternatively, the 1D $XXZ$ model can be mapped to a chain of spinless fermions by means of a Jordan-Wigner transformation (~\cite{takahashi,giamarchi}), where $J$ corresponds to the hopping to nearest neighbors and $J \Delta$ to a density-density interaction.}. This model can be exactly solved by means of the Bethe ansatz~\cite{takahashi,giamarchi,Sutherland}, and possesses several symmetries. Namely, it features a continuous $U(1)$ symmetry due to the conservation of the total magnetization in the $z$ direction for any $\Delta$ and an additional $SU(2)$ symmetry at $\Delta = \pm 1$ due to the conservation of the total magnetization along the $x$ and $y$ directions. Furthermore, the Hamiltonian is invariant under transformations $\sigma_{i}^{z}\to -\sigma_{i}^{z}$, thus having $\mathbb{Z}_{2}$ symmetry.

The model presents three different phases. First, for $\Delta<-1$ the ground state consists of a fully polarized configuration along the $z$ direction, i.e. it corresponds to a ferromagnetic state. In the intermediate regime $-1 < \Delta < 1$ the system is in a gapless phase, which can be shown to correspond to a Luttinger liquid in the continuum limit~\cite{giamarchi}. Finally, for $\Delta > 1$, the ground state corresponds to an antiferromagnetic configuration. The ferromagnetic and gapless states are separated by a first-order QPT at $\Delta = -1$, while the gapless and antiferromagnetic states are separated by a (infinite-order) Kosterlitz-Thouless QPT at $\Delta=1$.

\subsection{Spin-$\frac{1}{2}$  $X Y$ Model}	
We now describe the anisotropic 1D $X Y$ Hamiltonian for spins $1 / 2$. It corresponds to $J_{x}^{i,j}=\frac{1}{2}J(1+\gamma)$, $J_{y}^{i,j}=\frac{1}{2}J(1-\gamma)$, $B_z^i=B_z$, $J_{z}^{i,j}=B_x^i=B_y^i=0$ and is given by
\begin{equation} \label{hami_xy}
H=J\sum_{i}\left[\frac{1+\gamma}{2} \sigma_{i+1}^{x}\sigma_{i}^{x} + \frac{1-\gamma}{2}\sigma_{i+1}^{y}\sigma_{i}^{y}\right]+B_{z}\sum_{i}\sigma_{i}^{z}.
\end{equation}
Here $J > 0$ represents the exchange interaction between nearest neighbors, $\gamma > 0$ is the anisotropy parameter in the $X Y$ plane and $B_{z} > 0$ is the magnetic field along the $z$ direction. The limiting value  $\gamma = 1$ corresponds to the Ising model in a transverse magnetic field, which possesses a $\mathbb{Z}_2$ symmetry, and the limit $\gamma = 0$ is the isotropic $X Y$ model. In the thermodynamic limit $N \to \infty$, the anisotropic $X Y$ model can be exactly diagonalized by means of Jordan-Wigner and  Bogolyubov transformations~\cite{barouch1970pra,barouch1971pra}.

For the anisotropic case $0 < \gamma \leq 1$ the model belongs to the Ising universality class, and its phase diagram is determined by the ratio $\nu = 2B_z/J$.  When $\nu > 1$ the magnetic field dominates over the nearest-neighbor coupling, polarizing the spins along the $z$ direction. This corresponds to a paramagnetic state, with zero magnetization in the $xy$ plane. On the other hand, when $0 \leq \nu < 1$ the ground state of the system corresponds to a ferromagnetic configuration with polarization along the $xy$ plane. These phases are separated by a second-order QPT at the critical point $\nu = 1$. Finally, for the isotropic case $\gamma = 0$, a QPT is observed between gapless ($\nu < 2$) and ferromagnetic ($\nu > 2$) phases. We will only focus on the anisotropic model to illustrate the behavior of time correlations and LGI at criticality.


\section{Single-site two-time correlations} \label{section_time_correl}
Now we discuss how single-site two-time correlations (STC) can indicate different types of quantum phase transitions. We consider the symmetrized temporal correlation $C(t)$ for a single-site operator $A$, given by

\begin{equation}
C(t)=\frac{1}{2}\braGS \lbrace A(t),A(0)\rbrace \ketGS,
\end{equation}	
with $\ketGS$ the ground state of the time-independent Hamiltonian of interest $H$, $\lbrace .,.\rbrace$ the anticommutator between two operators, and $A (t)$ the operator at time $t$,
\begin{equation}
A(t)=e^{iHt}A(0)e^{-iHt}.
\end{equation}
For simplicity, we consider that $A(0)$ corresponds to one of the Pauli operators of a particular site $k$ $\left[ A(0) = \sigma_k^{\mu}\right.$, $\mu=x,y,z\left.\right]$. First note that the time correlations can be rewritten as
\begin{align} \label{time_correl}
\begin{split}
C(t)=Re\left[e^{iE_{0}t}\braGS A(0)e^{-iHt}A(0)\ketGS\right],
\end{split}
\end{align}
with $E_0$ the ground-state energy. Thus the correlation $ C (t) $ is a real quantity, which is fundamental for our subsequent analysis. As shown in Appendix~\ref{proof} for Hamiltonian~\eqref{Hami_general}, the STC of Eq.~\eqref{time_correl} and their derivatives constitute appropriate quantities to determine the location of finite-order QPTs.  In particular we obtain that the $(p-1)$th derivative of the STC is a function of $E_0$ and its first $p$ derivatives, so it can be written in the form
\begin{equation} \label{p_der_time_correl_final_prop_main}
\frac{\partial^{p-1} C(t)}{\partial \lambda^{p-1}} =\mathcal{F}\left(E_{0},\frac{\partial E_0}{\partial \lambda},\ldots,\frac{\partial^p E_0}{\partial \lambda^p},t\right),
\end{equation}
where $\lambda$ can be any Hamiltonian parameter, such as $\Delta$ for the $X X Z$ model and $\nu$ for the $X Y$ model. This means that, in general, a $p$th order QPT, which corresponds to a discontinuity or divergence of the $p$th derivative of the ground-state energy with respect to some Hamiltonian parameter, can be identified by the $(p-1)$th derivative of the STC with respect to the same parameter. Thus, a first-order QPT should in principle be identified by a discontinuity of the time correlations $C(t)$, at any time $Jt > 0$, as a function of the parameter driving the transition. Similarly, a second-order QPT should be recognized by a discontinuity or divergence of the first derivative of the time correlations with respect to the driving parameter. Note that this result is similar to the observation of finite-order QPTs by measures of bipartite entanglement~\cite{wu2004prl}. Here, however, we are able to determine transitions by looking at a purely local (single-site) quantity.

\begin{figure}
\begin{center}
\includegraphics[scale=0.43]{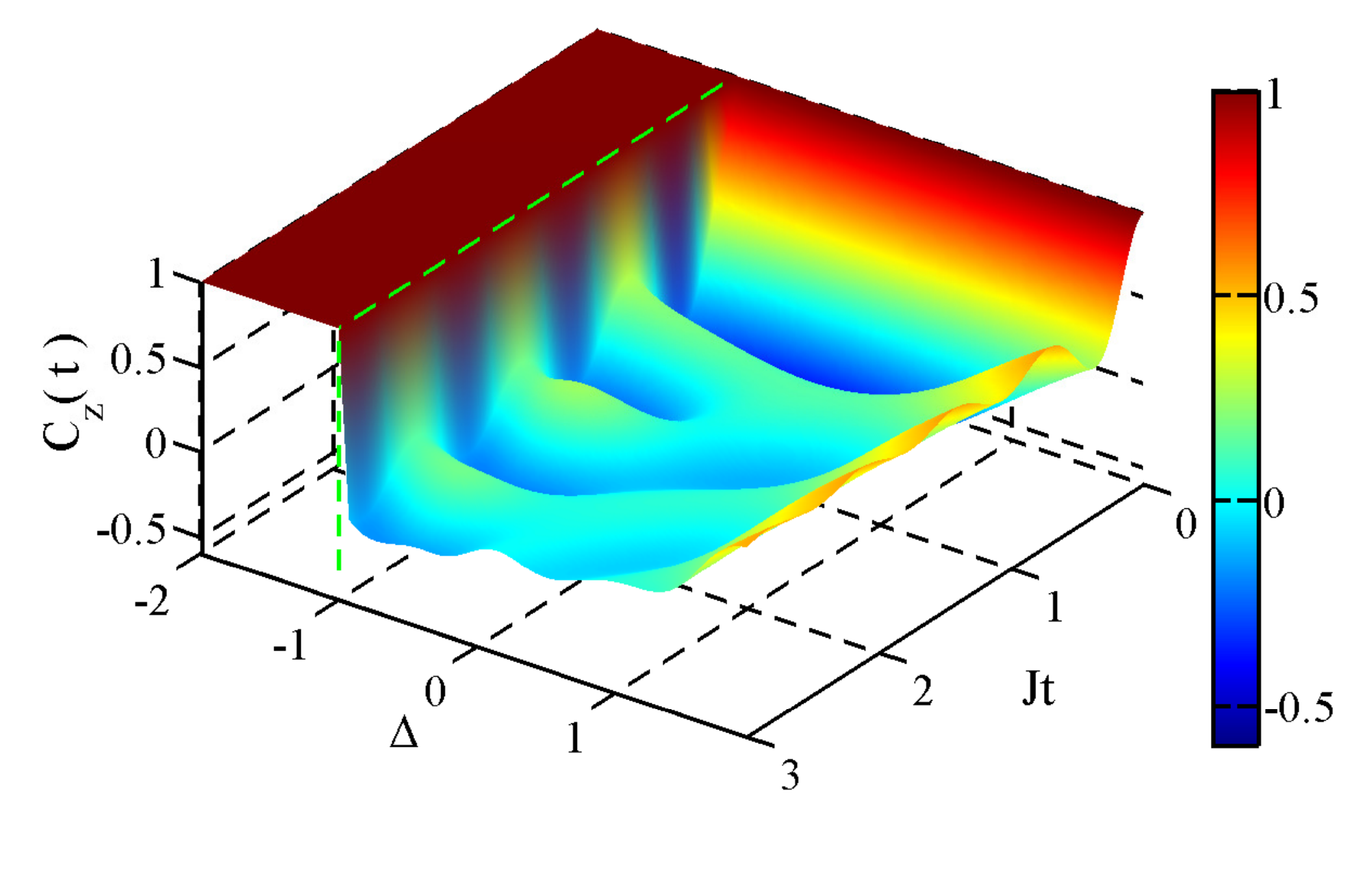}
\caption{\label{correl_z_xxz_3D} STC along the $z$ direction of the $X X Z$ model, as a function $J t$  and anisotropy parameter $\Delta$. The dashed green lines indicate the critical point $\Delta=-1$.}
\end{center}
\end{figure}

To provide stronger evidence that this is in fact the case, we calculate the time correlations for the $X X Z$ and anisotropic $X Y$ models, and examine their behavior at the corresponding quantum critical points. Even though both models are exactly solvable, obtaining their physical properties is a very challenging task. For example, for zero temperature exact time correlations are only known for $\Delta = 0$ in the $X X Z$ model (equivalent to the limit $\gamma = 0$ and $B_z = 0$ in the $X Y$ model) and for $A(0) = \sigma^z$~\cite{katsura1970physica}. Calculations based on a mean-field approach fail to reproduce the time correlations correctly. Furthermore, exact diagonalization methods are restricted to small lattices. Thus to obtain quantitatively-correct results for much longer systems we perform numerical simulations based on tensor-network algorithms. Namely, we first obtain the ground state of both models for several parameters by means of the density matrix renormalization group algorithm~\cite{white1992prl}, using a matrix product state description~\cite{schollwock2011ann}. Subsequently we simulate the time evolution described in Eq.~\eqref{time_correl} by means of the time evolving block decimation~\cite{vidal2004prl}. These methods allow us to carry out our simulations efficiently, for lattices of several sites. In particular, we consider systems of $N = 100$ spins (unless stated otherwise) with open boundary conditions, described by matrix product states with bond dimensions of up to $\chi = 400$. Our implementation of the algorithms is based on the open-source Tensor Network Theory (TNT) library~\cite{tnt}.

\subsection{First-order QPT}
\begin{figure}
\begin{center}
\includegraphics[scale=1]{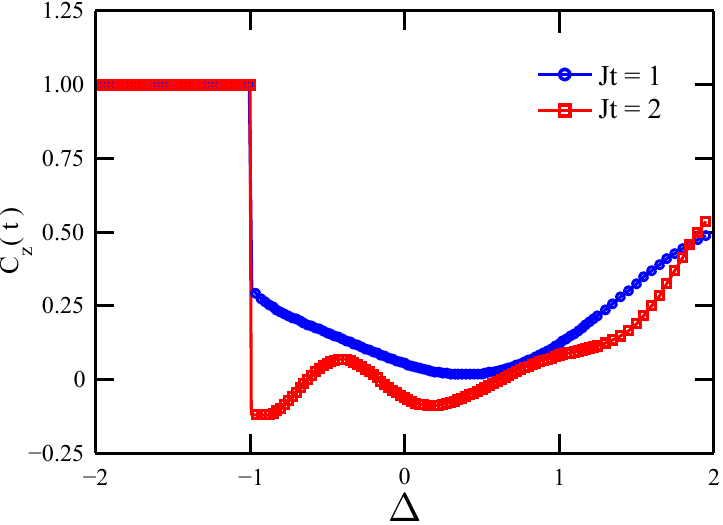}
\caption{\label{correl_z_xxz_2D} STC along the $z$ direction of the $X X Z$ model, as a function of $\Delta$, for  $Jt = 1$ and $Jt = 2$.}
\end{center}
\end{figure}

We start by observing the STC for the $X X Z$ model, and focus on the transition between ferromagnetic and gapless states at $\Delta = -1$. All the results to be presented are in a time scale from $0$ to $3$ in units of $1/J$. Since recent experiments on non-equilibrium spin models implemented in ultracold-atom quantum simulators have been performed for similar ($J/\hbar = 2\pi\times 8.6Hz$)~\cite{Fukuhara}  and even longer time scales ($J/\hbar = 14.1Hz$)~\cite{Hildprl}, the effects we will present are in a time scale perfectly observable with current technology. In Fig.~\ref{correl_z_xxz_3D} we show the correlations $C_z(t)$ $\left[\right.$i.e. for $A(0) = \sigma_k^z$$\left.\right]$, evaluated at site $k = 50$, as a function of $\Delta$ and $t$. Additionally, in Fig.~\ref{correl_z_xxz_2D} we plot the corresponding correlations at times $Jt = 1$ and $Jt = 2$ as a function of $\Delta$.

\begin{figure}[b]
\begin{center}
\includegraphics[scale=0.4]{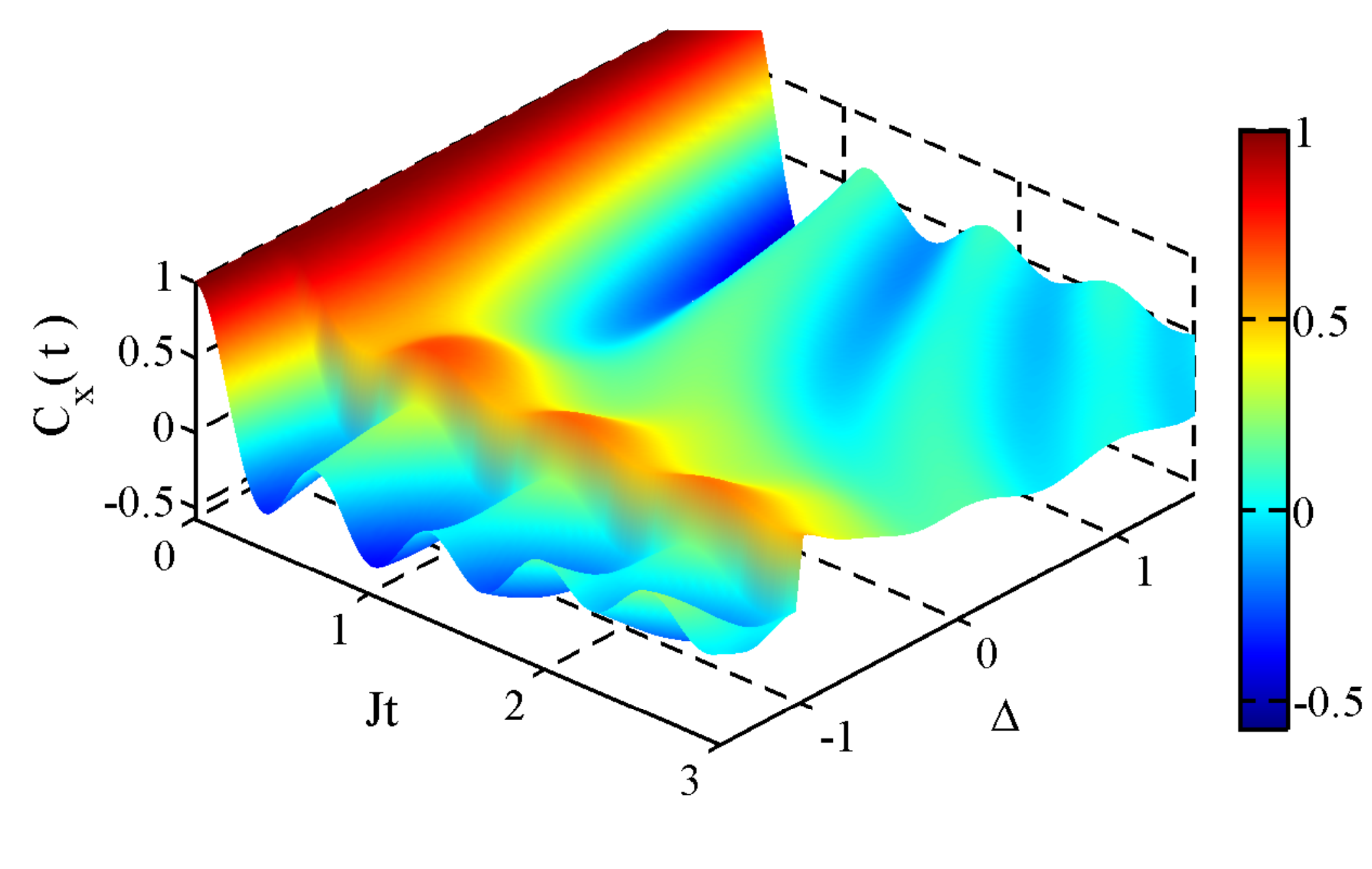}
\caption{\label{correl_x_xxz_3D} STC along the $x$ direction of the $XXZ$ model, as a function of  $J t$ and anisotropy parameter $\Delta$.}
\end{center}
\end{figure}

First, note that since the ground state of the system is ferromagnetic for $\Delta < -1$, so $\sigma_k^z\ketGS=\pm\ketGS$ in this regime, with the sign depending on the direction of polarization along the $z$ axis. Furthermore, this state remains unchanged under magnetization-conserving time evolution, such as that of the $XXZ$ model. Thus the time correlations remain constant, with value $C_z(t)=1$. For $\Delta>-1$ this is no longer the case. Since in this regime the states $\sigma_k^z\ketGS$ are not fully polarized, they are strongly affected by time evolution. More importantly, when the system crosses the quantum critical point $\Delta=-1$ and enters the gapless state, the correlations exhibit a discontinuous jump to values $C_z(t)<1$ at any finite time $Jt>0$, as depicted in Figs.~\ref{correl_z_xxz_3D} and~\ref{correl_z_xxz_2D}.

\begin{figure}
\begin{center}
\includegraphics[scale=1.1]{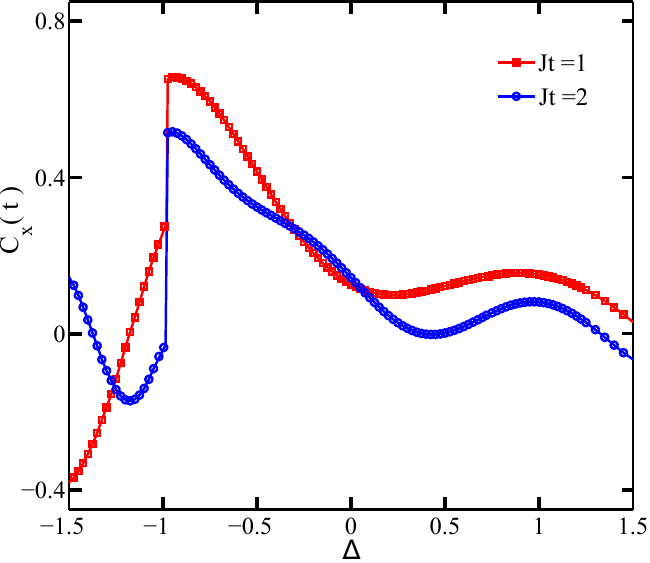}
\caption{\label{correl_x_xxz_2D} STC along the $x$ direction of the $XXZ$ model, as a function of $\Delta$, for $Jt=1$ and $Jt=2$.}
\end{center}
\end{figure}

\begin{figure}[b]
\begin{center}
\includegraphics[scale=0.4]{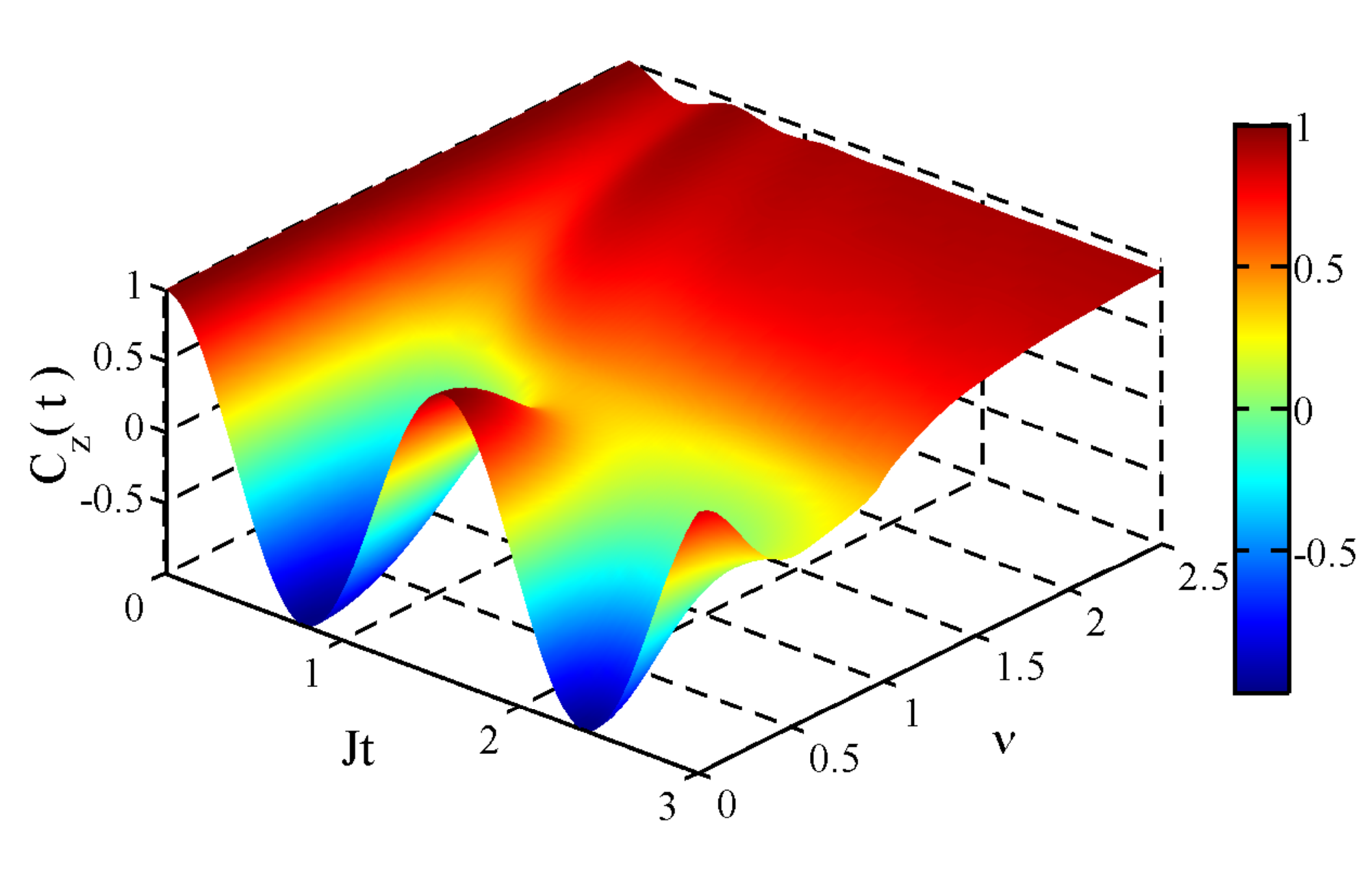}
\caption{\label{correl_z_xy_3D} STC along the $z$ direction of the anisotropic $XY$ model for $\gamma=1$ (Ising model), as a function of  $Jt$ and $\nu$. }
\end{center}
\end{figure}

A similar result is obtained when calculating the STC $C_x(t)$, i.e. with $A(0)=\sigma_k^x$, which give identical results to the correlations along the $y$ direction due to the symmetry of the Hamiltonian~\eqref{hami_xxz}. These are shown in Fig.~\ref{correl_x_xxz_3D} as a function of $\Delta$ and $Jt$, and in Fig.~\ref{correl_x_xxz_2D} for two specific times, namely $Jt=1$ and $Jt=2$. In contrast to $C_z(t)$, the correlations along the $x$ direction do not remain constant in the ferromagnetic regime $\Delta<-1$, since $\sigma_k^x$ flips the spin at site $k$ and thus induces dynamics on the system. However, the correlations $C_x(t)$ also show a discontinuity at $\Delta=-1$. Thus as expected from Eq.~\eqref{p_der_time_correl_final_prop_main}, the different STC indicate the first-order QPT of the $XXZ$ model by means of a discontinuity as a function of $\Delta$ at the quantum critical point.

Note that neither $C_z(t)$ nor $C_x(t)$, or any of their derivatives, indicate the existence of the Kosterlitz-Thouless QPT at $\Delta=1$,  given that it is of infinite order. However, we will observe in Section~\ref{lgi_kt} that it is possible to identify this transition by the maximization of the violation of LGI.
\subsection{Second-order QPT}

\begin{figure}
\begin{center}
\includegraphics[scale=0.4]{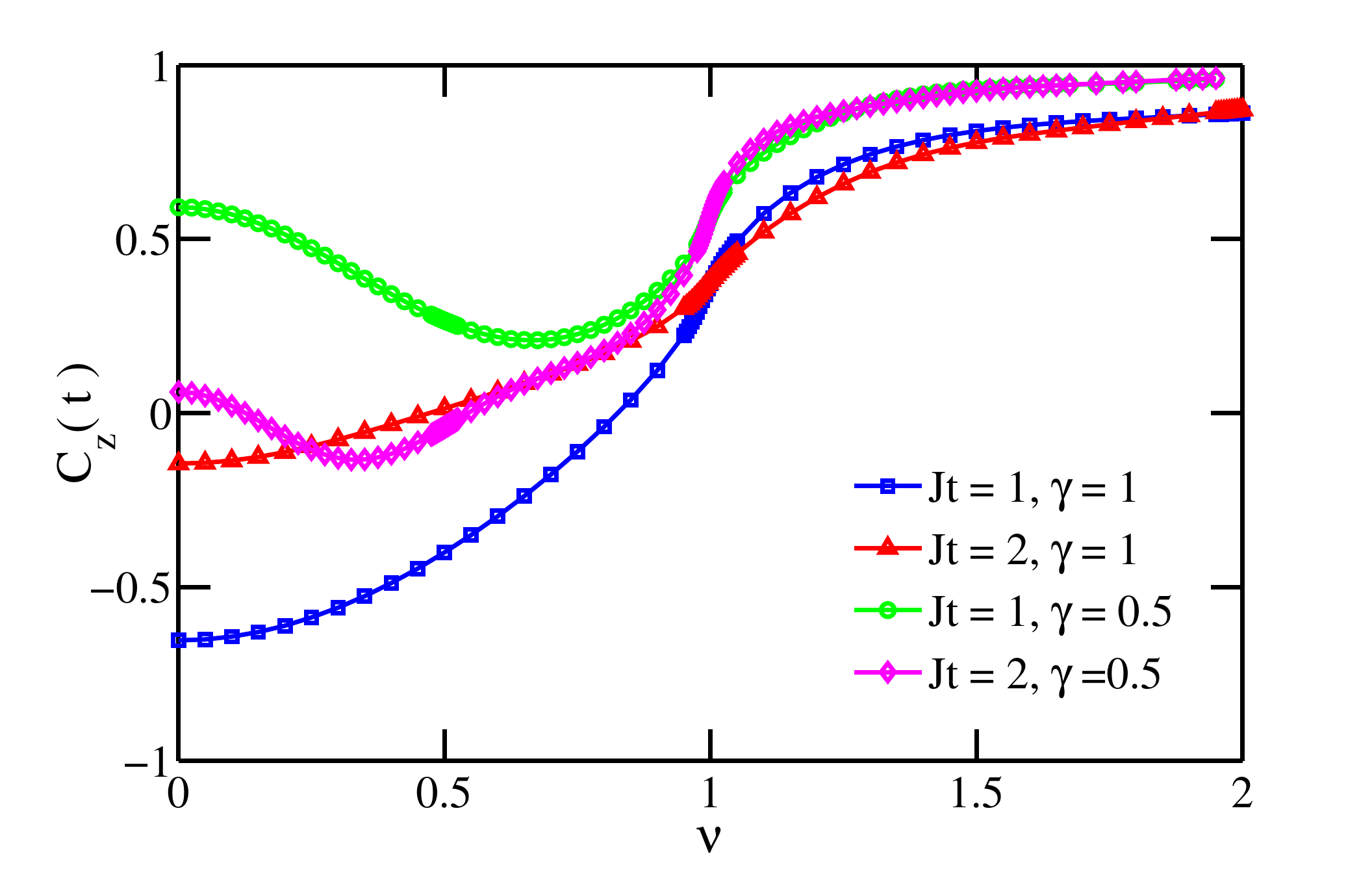}
\includegraphics[scale=0.4]{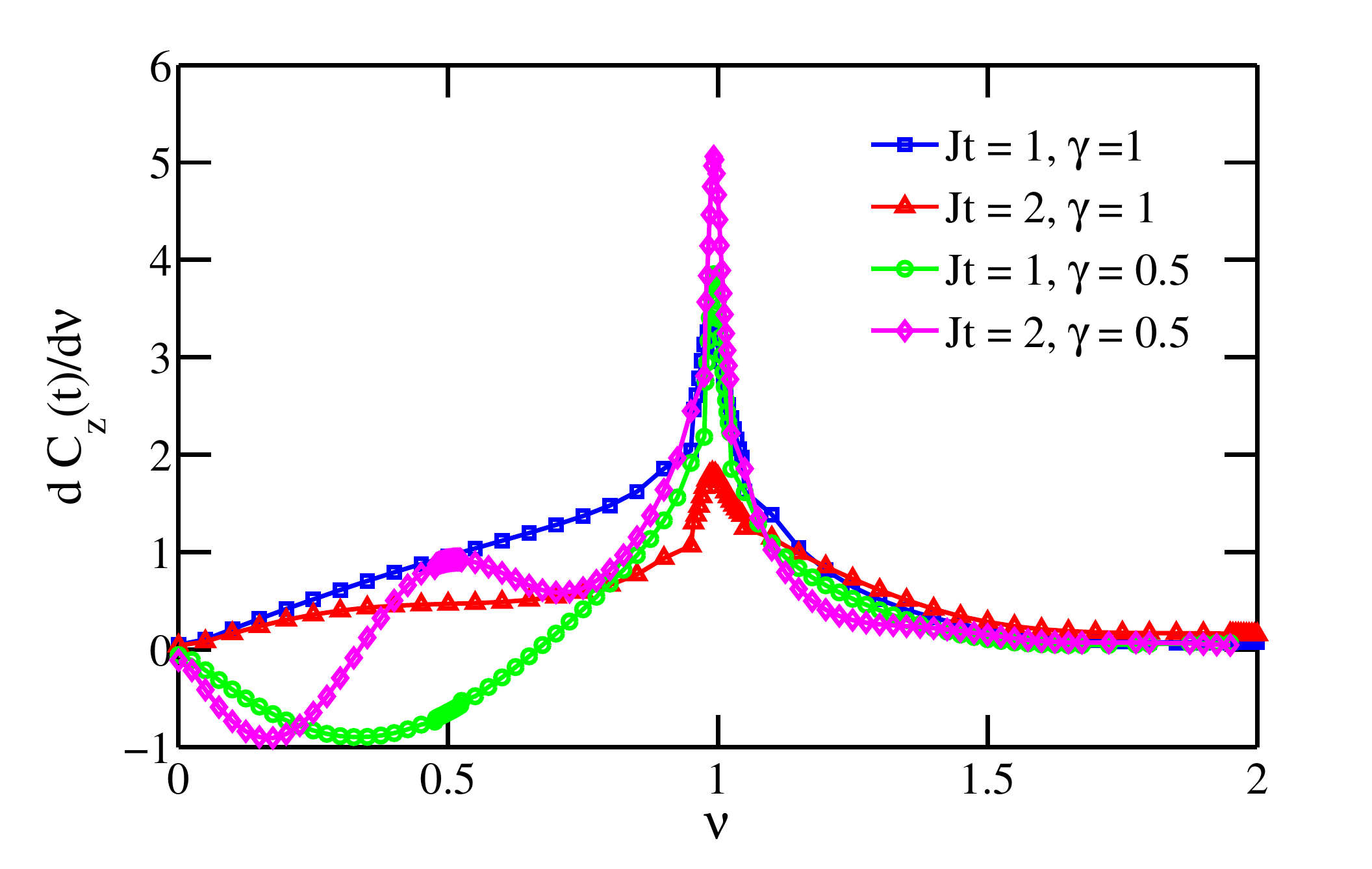}
\caption{\label{correl_z_xy_2D} STC (upper panel) and first derivative (lower panel) along the $z$ direction of the anisotropic $XY$ model ($\gamma=0.5,1$), as a function of $\nu$, for  $Jt=1$ and $Jt=2$.}
\end{center}
\end{figure}

Now we consider the transition between ferromagnetic and paramagnetic phases of the anisotropic $XY$ model. In particular, we illustrate the transition for two cases, namely the limit $\gamma=1$, which corresponds to the Ising model with a transverse magnetic field, and the intermediate case $\gamma=0.5$.  We verified that the STC along any direction $ \alpha = x, y, z $ give the same qualitative information regarding the QPT, which is in agreement with Eq.~\eqref{p_der_time_correl_final_prop_main}, so we focus on $ C_z (t) $ and do not show the results of the other directions.

In Fig.~\ref{correl_z_xy_3D} we show the $z$ time correlations as a function of $Jt$ and $\nu$, for $\gamma=1$; the results for $\gamma=0.5$ are qualitatively similar. In addition, we depict in the upper panel of Fig.~\ref{correl_z_xy_2D} the correlations for times $Jt=1$ and $Jt=2$ as a function of the magnetic field, for both $\gamma=1$ and $\gamma=0.5$. In contrast to the $XXZ$ case, here the correlations are continuous for the whole range of values of $\nu$ considered. However, the first derivative with respect to $\nu$ is not a well-behaved function. As exemplified in the lower panel of Fig.~\ref{correl_z_xy_2D} for two particular times, $\frac{dC_z(t)}{d\nu}$ shows a sharp maximum at the quantum critical point $\nu=1$. Thus, in accordance with Eq.~\eqref{p_der_time_correl_final_prop_main}, the second-order QPT of the model can be identified by means of a singularity in the first derivative of the local time correlations with respect to the Hamiltonian parameter which drives the transition, which in this case is $\nu$.
\section{Leggett-Garg inequalities} \label{sect_lgi}

Since the birth of quantum mechanics, its non-deterministic nature and nonlocal structure have motivated many theoretical debates that have recently moved to the experimental field. In particular, Bell inequalities establish a natural border to the spatial quantum correlations in separate systems. Leggett and Garg~\cite{lgi_original} in 1985 showed that the temporal correlations obey similar inequalities.

\begin{figure}
\begin{center}
\includegraphics[scale=0.8]{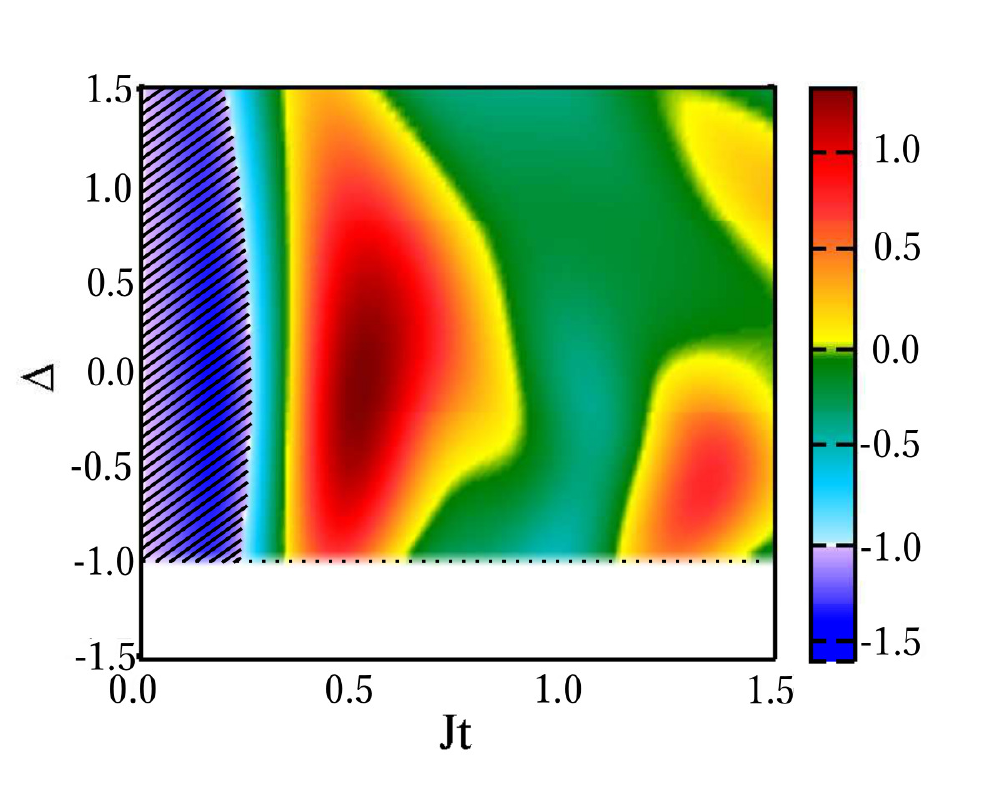}
\includegraphics[scale=0.77]{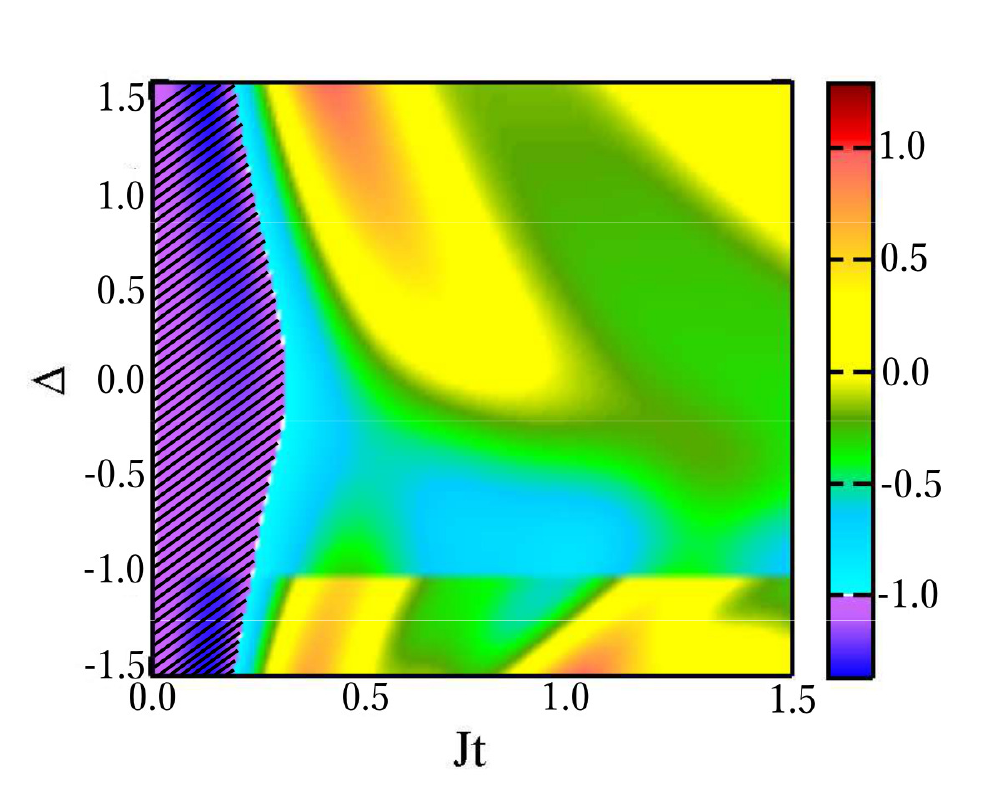}
\caption{\label{lgi_z_and_x_xxz_3D} Leggett-Garg functions for the $XXZ$ model. Upper panel: $K_{-}^{z}(t)$. The regions with diagonal lines corresponds to the regime of anisotropies $\Delta$ and  $Jt$ in which the $x$ Leggett-Garg inequalities are violated. The white region below $\Delta=-1$ indicates that there the time correlations remain constant. Lower panel: $K_{-}^{x}(t)$. The  regions with diagonal lines region corresponds to the regime of anisotropies $\Delta$ and  $Jt$ in which the $x$ Leggett-Garg inequalities are violated.}
\end{center}
\end{figure}

In our intuitive view of the world, probabilities are due to our uncertainty about the state of a system, but they are not a fundamental description of it. For example, when we toss a coin to the air, it has probability one half of landing tails or heads. We also assume that if we had the precise knowledge of its position and momentum, and enough computational power, we would be able to determine on which side the coin will land. We do not think that the coin is in a superposition of states, such a Schr\"odinger's cat. This is known as macroscopic realism. In addition, we assume that making measurements on a system does not modify its present state, in the way projective quantum measurements do. This is referred to as noninvasive measurability. Based on these two principles Leggett and Garg obtained a set of inequalities, which is consistent with the macroscopic intuition. One form these LGI can take is

\begin{equation}\label{LGI}
C\left(t_{1},t_{3}\right)-C\left(t_{1},t_{2}\right)-C\left(t_{2},t_{3}\right)\geq -1,
\end{equation}
where $C\left(t_{i},t_{j}\right)=\frac{1}{2}\langle \lbrace Q(t_{i}),Q(t_{j})\rbrace\rangle$ is the two-time correlation of a dichotomic observable	$Q$ (with eigenvalues $q=\pm1$) between times $t_i$ and  $t_j$, and $t_1 < t_2 < t_3$. On the other hand, if the correlation functions $C\left(t_{i},t_{j}\right)$ are stationary, i.e., they only depend on the time difference $\tau=t_i -t_j$, then the Leggett-Garg inequality~\eqref{LGI} can be written as~\cite{huelga1995pra}
\begin{equation}\label{eq9}
K_{-}\left(\tau\right) \equiv C\left(2\tau \right) - 2C\left(\tau\right)\geq-1,
\end{equation}
which defines the Leggett-Garg functions $K_-(\tau)$ for time $\tau$. Just as with Bell inequalities, any system that violates inequality~\eqref{eq9} shows some behavior that is essentially nonclassical. This is why violations of LGI are used as a measure of quantumness~\cite{HuelgaPRL2015}. In the following we discuss different Leggett-Garg functions $K_{-}^{\alpha}(t)$, corresponding to measurements of spin components along the $\alpha$ direction, and see whether they can give information about the QPTs previously discussed.
\subsection{Finite-order QPTs}

\begin{figure}
\begin{center}
\includegraphics[scale=1.1]{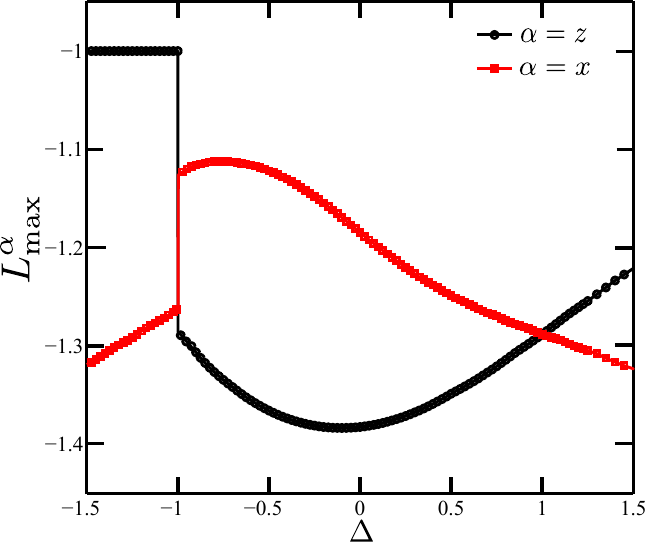}
\caption{\label{lgi_max_xxz_x_z} Maximum violation of Leggett-Garg inequalities for the $XXZ$ model, along both $z$ and $x$ directions.}
\end{center}
\end{figure}
We start by showing how the Leggett-Garg functions $K_{-}^{\alpha}(t)$ signal the finite-order QPTs discussed in Sec.~\ref{section_time_correl}. In Fig.~\ref{lgi_z_and_x_xxz_3D} we depict both $K_{-}^{z}(t)$ (upper panel) and $K_{-}^{x}(t)$ (lower panel) for the $X X Z$ model as a function of $\Delta$  and time. Regarding the results along the $\alpha=z$ direction, we first note that for $\Delta<-1$ the value of the Leggett-Garg function remains equal to $K_{-}^{z}(t) = -1$ for any time. Thus in the ferromagnetic phase the corresponding LGIs are never violated. This is clearly a direct consequence of the constant value of the time correlations previously discussed (see Figs.~\ref{correl_z_xxz_3D} and~\ref{correl_z_xxz_2D}), and manifests the classical nature of the ferromagnetic state when undisturbed. The situation is entirely different for $\Delta>-1$. Not only $K_{-}^{z}(t)$ does vary on time, but it indicates a violation of the Leggett-Garg inequalities for early times. As to the results along the $\alpha=x$ direction, all the values of $\Delta$ considered show a violation of the inequalities for early times. In addition, the violation lasts longer as $|\Delta|$ decreases.
\begin{figure}
\begin{center}
\includegraphics[scale=0.85]{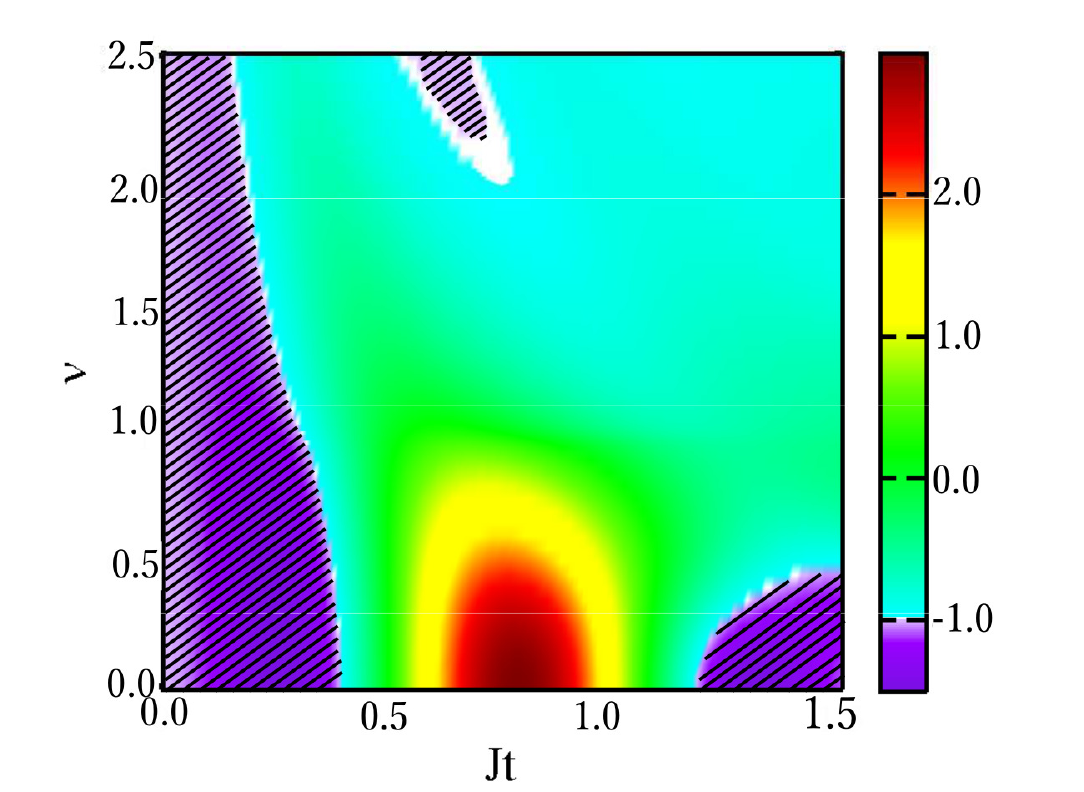}
\includegraphics[scale=0.85]{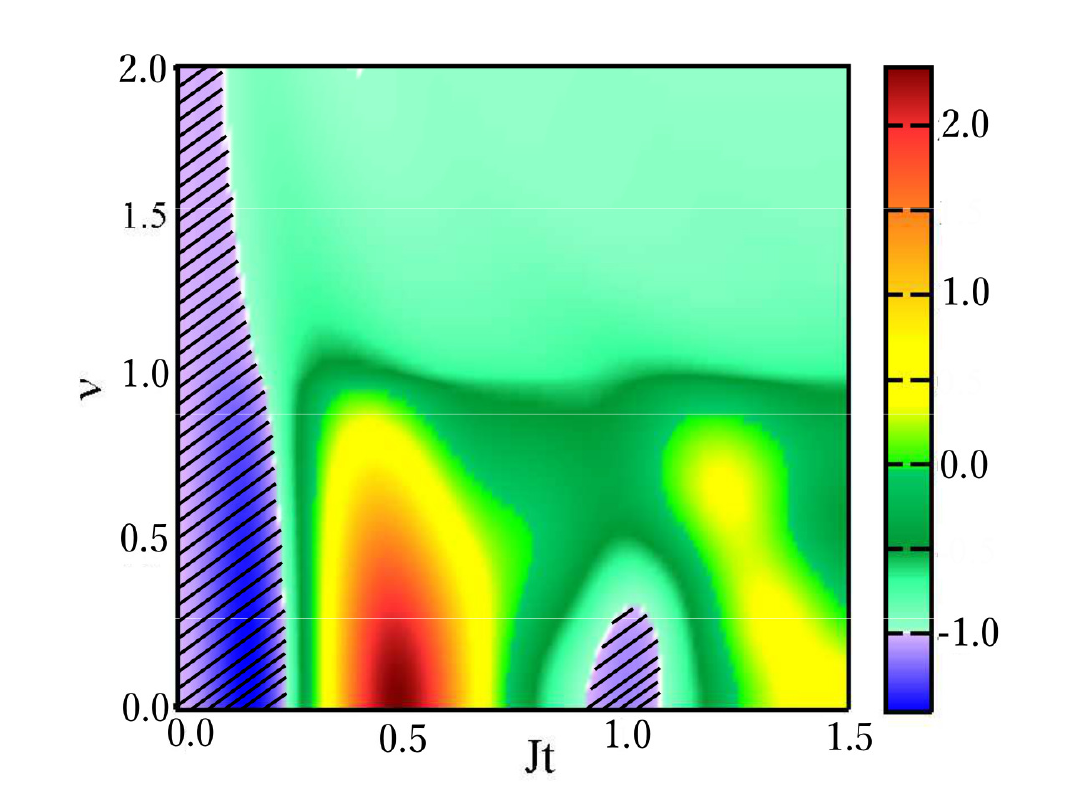}
\caption{\label{lgi_z_gamma_1_and_05_xy_3D} Leggett-Garg functions $K_{-}^{z}(t)$ for the anisotropic $XY$ model. Upper panel: $\gamma=1$. Lower panel: $\gamma=0.5$. The regions with diagonal lines corresponds to the regime of parameter  $\nu$ and  $Jt$ in which the $z$ Leggett-Garg inequalities are violated.}
\end{center}
\end{figure}
In Fig.~\ref{lgi_max_xxz_x_z} we plot the maximum value $L_{\text{max}}^{\alpha}$ we obtain for the violation of the Leggett-Garg inequalities as defined by
\begin{equation}
L_{\text{max}}^{\alpha}=\min_{t}K_{-}^{\alpha}(t),
\end{equation}
for both directions $\alpha=z,x$ as a function of $\Delta$. This clearly shows that similarly to time correlations, the first-order QPT of the $XXZ$ model can be identified by a discontinuity of the $L_{\text{max}}^{\alpha}$ function at the critical point. Also, the maximal violation occurs along the $z$ direction, close to the noninteracting limit $\Delta=0$. In contrast, for the magnetically-ordered phases, the maximal violation occurs along the $x$ direction.

The Leggett-Garg functions also help determine the second-order QPT of the anisotropic $XY$ model. In Fig.~\ref{lgi_z_gamma_1_and_05_xy_3D} we show $K_{-}^{z}(t)$ for several values of $\nu$ as a function of time, for $\gamma=1$ (upper panel) and $\gamma=0.5$ (lower panel). Notably, for all the values of $\nu$ considered, the system features the violation of the inequalities. Initially, for $Jt < 0.5$, the violation of the inequalities lasts longer as $\nu$ decreases. Interestingly, for longer times, revivals of the violations are seen for low values of $\nu$. Thus weak magnetic fields favor the observation of the violation of the Leggett-Garg inequalities along $z$ direction. 

\begin{figure}
\begin{center}
\includegraphics[scale=1.3]{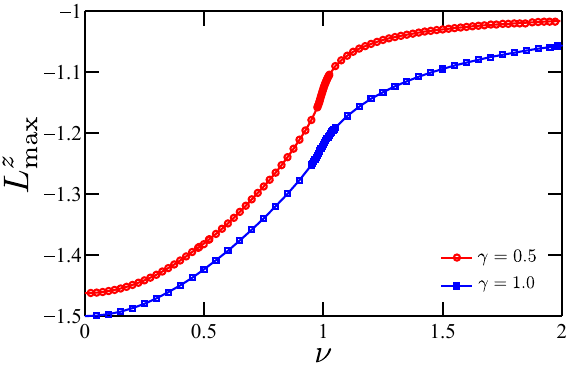}
\includegraphics[scale=1.3]{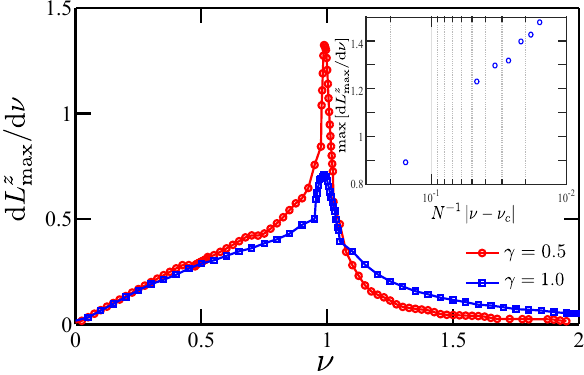}
\caption{\label{lgi_z_max_xy} Upper panel: Maximum violation of Leggett-Garg inequalities for the anisotropic $X Y$ model ($\gamma = 1,0.5$) along the $z$ direction, as a function of $\nu$. Lower panel: First derivative with respect to $\nu$.  Inset:  scaling of  maximum of the derivative with respect to $\nu$} 
\end{center}
\end{figure}

Just as the time correlations, the Leggett-Garg functions $K_{-}^{z}(t)$ and the maximal violation functions $L_{\text{max}}^{z}$ (see upper panel of Fig.~\ref{lgi_z_max_xy}) are continuous in the whole parameter regime. However, their first derivative tends to diverge at the quantum critical point $\nu=1$ as the size of the system increases (see inset lower panel of Fig.~\ref{lgi_z_max_xy}). This is shown in the lower panel of Fig.~\ref{lgi_z_max_xy} for $L_{\text{max}}^{z}$ and both $\gamma = 1$ and $\gamma = 0.5$. As expected, the behavior of the time correlations is translated to the Leggett-Garg functions, and they are able to signal the second-order QPT of the anisotropic $XY$ model by means of a singularity in their first derivative.

\subsection{Infinite-order QPT of the XXZ model} \label{lgi_kt}

We have observed that finite-order QPTs can in principle be determined by means of a singular behavior of local unequal-time correlations and Leggett-Garg functions, or of their derivatives. However, this form is not suitable to identify infinite-order transitions. In fact, the results shown so far do not feature any singular property at the quantum critical point $\Delta=1$ of the Kosterlitz-Thouless transition of the one-dimensional $XXZ$ model. However, it is possible to locate this transition from Leggett-Garg functions, as we discuss in the present Section. This is very similar to the observation of the transition from Bell inequalities~\cite{justino2012pra}, with the notable difference that here we actually have violation of the respective inequalities, and thus we can perceive the quantumness of the system.

\begin{figure}
\begin{center}
\includegraphics[scale=1.1]{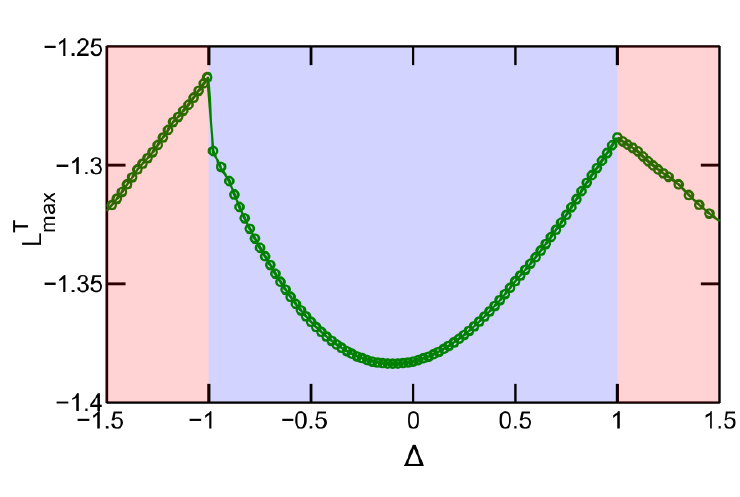}
\caption{\label{lgi_max_xxz_tot} Total maximum violation of Leggett-Garg inequalities for the $X X Z$ model as a function of $\Delta$. The light-red zones indicate the regimes in which the maximal violations comes from inequalities along the $x$ direction, while the light-blue zone in between shows the regime in which the maximal violations occur along the $z$ direction; see Fig.~\ref{lgi_max_xxz_x_z}.}
\end{center}
\end{figure}

The first point to note is that to actually establish that a violation of the inequalities exists, and also when the maximal violation occurs, we must consider all the possible directions $\alpha$ of evaluation of time correlations. For the $X X Z$ model, this corresponds to $\alpha = z$ and $\alpha=x$, the latter giving the same results as for $\alpha = y$ due to the symmetry of the Hamiltonian. For this we define the function
\begin{equation}
L_{\text{max}}^{\text{T}}=\max_{\alpha=z,x}L_{\text{max}}^{\alpha},
\end{equation}
which maximizes over all times and directions the violation of the Leggett-Garg inequalities. We show $L_{\text{max}}^{\text{T}}$ as a function of $\Delta$ in Fig.~\ref{lgi_max_xxz_tot}; note that it indicates the first-order QPT at $\Delta=-1$ by means of a discontinuity.

The second point to note, responsible for the observation of the Kosterlitz-Thouless transition by means of Bell inequalities, is that at the isotropic point of the Hamiltonian there is a change in the largest type of spatial correlation~\cite{justino2012pra}. Namely, for $|\Delta|>1$ and spins separated by $r$ lattice sites, $|\langle\sigma_i^x\sigma_{i+r}^x\rangle|\leq|\langle\sigma_i^z\sigma_{i+r}^z\rangle|$, while for $-1<\Delta<1$ we have that $|\langle\sigma_i^x\sigma_{i+r}^x\rangle|>|\langle\sigma_i^z\sigma_{i+r}^z\rangle|$. As expected from the discussion of Appendix~\ref{proof}, this behavior is translated to the local time correlations and related functions. In fact, as seen in Fig.~\ref{lgi_max_xxz_x_z}, $L_{\text{max}}^{x}<L_{\text{max}}^{z}$ for the ordered phases, while $L_{\text{max}}^{x}>L_{\text{max}}^{z}$ for the gapless regime. As shown in Fig.~\ref{lgi_max_xxz_tot}, a sharp local maximum appears at $\Delta=1$ in the $L_{\text{max}}^{\text{T}}$ function, and a singularity of its first derivative results. Thus, by means of a function characterizing the total maximal violation of Leggett-Garg inequalities, we are able to locate the infinite-order Kosterlitz-Thouless transition of the $XXZ$ model. It would be interesting to observe whether Kosterlitz-Thouless transitions for other quantum systems can be identified in this form.

\section{Conclusions} \label{conclu}

In the present paper we have discussed whether single-site time correlations and Leggett-Garg inequalities allow the identification of QPTs in many-body quantum systems. By means of efficient matrix product simulations and analytical arguments, we have answered this question in the affirmative for different spin-$1/2$ models, for both finite- and infinite-order QPTs. Thus we have shown that QPTs can be detected by purely-local measurements.

Initially, by means of a first-order approximation for a general spin-$1/2$ Hamiltonian, we argued that a $p$th order QPT can be located by a singular behavior of the $(p-1)$th derivative of the local time correlations at the quantum critical point. Thus, these correlations indicate quantum criticality in a form similar to different measures of bipartite entanglement~\cite{wu2004prl}. Furthermore, this behavior is directly transferred to the corresponding Leggett-Garg functions.

To support this general result, we calculated several time correlations for large one-dimensional $XXZ$ and anisotropic $XY$ spin systems, using the density matrix renormalization group and time evolving block decimation methods. In particular, we showed that the first-order ferromagnetic-gapless QPT of the $XXZ$ model is manifested as a discontinuity of the correlations at $\Delta=-1$, along any possible direction and for any finite time. Subsequently we showed that the second-order paramagnetic-ferromagnetic QPT of the anisotropic $XY$ model is observed by means of a divergence of the first derivative of the correlations with respect to the magnetic field at $\nu=1$.

We also showed that the Leggett-Garg functions can help identify finite-order QPTs in a similar fashion. More importantly, we found that at least for one direction, the Leggett-Garg inequalities are violated for early times and the whole regime of parameters considered, in contrast to Bell inequalities~\cite{justino2012pra}. Furthermore, the maximization of this violation allowed us to identify the infinite-order Kosterlitz-Thouless QPT of the $XXZ$ model at $\Delta=1$, which was not possible from the separate observation of time correlations along each direction. Given the large amount of materials described by the test-bed models discussed in our work~\cite{aeppli,blundell_magnetism}, and the seminal advances on their implementation in quantum simulators~\cite{georgescu2014rmp}, we expect that our results extend the range of systems in which the violation of Leggett-Garg inequalities can be observed experimentally~\cite{emary2014rpp}.

For future research, it would be interesting to analyze whether local time correlations and Leggett-Garg inequalities can identify the existence of different-order nonequilibrium quantum phase transitions~\cite{prosen2008prl,prosen2010exact,prosen2010prl,bastidas2012prl,Acevedo2014PRL,bastidas2012pra,we,ajisaka2014njp}. Similar analysis on Bell-Leggett-Garg inequalities~\cite{bell_leggett_garg_2015} could lead to important insights into the relation between measures of quantumness and quantum criticality.

\begin{acknowledgments}
We acknowledge Sarah Al-Assam and the TNT Library Development Team for providing the codes for the simulations carried out during our work. C.T. acknowledges financial support from the Spanish MINECO under contracts MAT2011-22997 and MAT2014-53119-C2-1-R. F.J.G.R  acknowledges financial support from Proyectos Semilla-Facultad de Ciencias at Universidad de los Andes (2015). F.J.G.R, F.J.R., and L.Q. acknowledge financial support from Facultad de Ciencias at UniAndes-2015 project  \textquotedblleft Quantum control of nonequilibrium hybrid systems-Part II\textquotedblright.
\end{acknowledgments}

\appendix
		
\section{Proof of the relations between two-time correlations and ground-state energy} \label{proof}
		
In this section we show how STC can indicate finite-order QPTs of spin-$1/2$ systems of any dimensionality or coupling range. Thus we consider a Hamiltonian of the form
\begin{align} \label{hami_appendix}
\begin{split}
H=&\sum_{\alpha}\sum_{i,j} J_{\alpha}^{i,j}\sigma_i^{\alpha}\sigma_j^{\alpha}+\sum_{\alpha}\sum_{i}B_{\alpha}^{i}\sigma_i^{\alpha}\\
=&\sum_{\alpha}\sum_{i,j}J_{\alpha}^{i,j}\frac{\partial H}{\partial J_{\alpha}^{i,j}}+\sum_{\alpha}\sum_{i}B_{\alpha}^i\frac{\partial H}{\partial B_{\alpha}^i}.
\end{split}
\end{align}
As shown in Eq.~\eqref{time_correl}, the time correlations of a single-site operator $A$ are given by
\begin{align}
C(t)=Re\left[e^{iE_0t}\braGS A(0)e^{-iHt}A(0)\ketGS\right]
\end{align}
Now we expand the time evolution operator as
\begin{equation}
e^{-iHt}=\sum_{l=0}^{\infty} \frac{\left(-it\right)^{l}}{l!}H^{l}
\end{equation}
So the product $A(0)e^{-iHt}A(0)$ can be written as
\begin{align}
A(0)&e^{-iHt}A(0)=\sum_{l=0}^{\infty} \frac{\left(-it\right)^{l}}{l!}A(0)H^{l}A(0)\notag\\
&=\sum_{l=0}^{\infty} \frac{\left(-it\right)^{l}}{l!}\left(A(0)HA(0)\right)^{l}=e^{-it A(0)HA(0)}
\end{align}
where we have used $A(0)A(0)=\left(\sigma_k^{\mu}\right)^2=\mathcal{I}$ repeatedly in the first line to obtain the second equality ($\mu=x,y,z$).

Using the explicit form~\eqref{hami_appendix} of the Hamiltonian, we obtain after straightforward algebra that
\begin{align}
A(0)HA(0)&=H - f_{k}^{\mu}.
\end{align}
where the second term explicitly depends on site $k$ where the correlations are calculated, namely
\begin{align}
f_{k}^{\mu}=2\sum_{\alpha\neq\mu}\left(\sum_{j}J_{\alpha}^{j,k}\sigma_{k}^{\alpha}\sigma_{j}^{\alpha} +B_{\alpha}^{k}\sigma_{k}^{\alpha}\right)
\end{align}
Therefore, the STC is given by
\begin{equation} \label{correlation}
C(t)=Re\left[e^{i E_{0}t}\braGS e^{-it\left(H-f_{k}^{\mu}\right)}\ketGS\right].
\end{equation}
To easily observe the relation between QPTs and time correlations, we restrict to first order in the exponential within the expectation value of Eq.~\eqref{correlation}. In this case, we have
\begin{align}
C(t)&\approx \cos\left(E_{0}t\right)+\sin\left(E_{0}t\right)\left[E_{0}t -t \braGS f_{k}\ketGS \right]
\end{align}
Now we rewrite $f_k^{\mu}$ in the form
\begin{align}
f_{k}^{\mu}&=2\sum_{\alpha\neq\mu}\left( \sum_j J_{\alpha}^{j,k}\frac{\partial H}{\partial J^{j,k}_{\alpha}} + B_{\alpha}^k \frac{\partial H}{\partial B_{\alpha}^k}\right).
\end{align}
Finally, using the Hellmann-Feynman relations
\begin{equation}
\left\langle\psi_{0}\left|\frac{\partial H}{\partial\lambda}\right|\psi_{0}\right\rangle=\frac{\partial E_0}{\partial\lambda},
\end{equation}
with $\lambda$ any parameter of the Hamiltonian~\cite{griffiths}, we obtain that up to first order in the expansion of the time evolution operator, the time correlations are given by
\begin{align} \label{time_correl_final}
C(t)\approx & \cos\left(E_{0}t\right)+\sin\left(E_{0}t\right)\bigg[E_{0}t\notag\\
&-2t\sum_{\alpha\neq\mu}\left(\sum_{j}J_{\alpha}^{j,k}\frac{\partial E_{0}}{\partial J_{\alpha}^{j,k}} +B_{\alpha}^{k}\frac{\partial E_{0}}{\partial B_{\alpha}^{k}}\right)\biggr]
\end{align}
Thus we have obtained that the STC are proportional (apart from a structureless term $E_{0}t$)  to the first derivatives of the ground-state energy of the system, which show a discontinuity at the critical point of a first-order QPT~\cite{wu2004prl}. This means that first-order QPTs are directly identified by discontinuities of the STC as a function of Hamiltonian parameters.\\
\\
Now we consider the first derivative of Eq.~\eqref{time_correl_final} with respect to some Hamiltonian parameter, e.g. $J_{\beta}^{m,n}$. We obtain
\begin{align} \label{1der_time_correl_final}
\frac{\partial C(t)}{\partial J_{\beta}^{m,n}}&\approx t\cos\left(E_{0}t\right)\frac{\partial E_{0}}{\partial J_{\beta}^{m,n}}\notag\\
&\bigg[E_{0}t-2t\sum_{\alpha\neq\mu}\left(\sum_{j}J_{\alpha}^{j,k}\frac{\partial E_{0}}{\partial J_{\alpha}^{j,k}} +B_{\alpha}^{k}\frac{\partial E_{0}}{\partial B_{\alpha}^{k}}\right)\biggr]\notag\\
&-2t\sin\left(E_{0}t\right)\sum_{\alpha\neq\mu}\Biggr[\sum_{j}\Biggr(\delta_{\alpha,\beta}\delta_{m,j}\delta_{n,k}\frac{\partial E_{0}}{\partial J_{\alpha}^{j,k}}\notag\\
&+J_{\alpha}^{j,k}\frac{\partial^{2}E_{0}}{\partial J_{\beta}^{m,n}\partial J_{\alpha}^{j,k}}\Biggr)+B_{\alpha}^{k}\frac{\partial^{2}E_{0}}{\partial J_{\beta}^{m,n}\partial B_{\alpha}^{k}}\Biggr]
\end{align}
This means that the first derivative of the STC with respect to a Hamiltonian parameter is proportional to the second derivative of the ground-state energy with respect to the same parameter,
\begin{equation} \label{1der_time_correl_final_prop}
\frac{\partial C(t)}{\partial J_{\beta}^{m,n}}\propto t\frac{\partial^2 E_0}{\partial (J^{m,n}_{\beta})^2}\quad\text{and}\quad\frac{\partial C(t)}{\partial J_{\beta}^{m,n}}\propto E_0t.
\end{equation}
As a result of the first proportionality relation of Eq.~\eqref{1der_time_correl_final_prop}, the derivatives of unequal-time correlations indicate second-order QPTs by means of a discontinuity or divergence at the corresponding quantum critical points.

The previous results indicate that, in general, a finite-order QPT can be identified by the properties of the STC of the system. Namely, given that
\begin{align} \label{p_der_time_correl_final_prop}
\begin{split}
&\frac{\partial^{p-1} C(t)}{\partial (J_{\beta}^{m,n})^{p-1}}\propto t\frac{\partial^p E_0}{\partial (J^{m,n}_{\beta})^p},\\
\text{and}\quad&\frac{\partial^{p-1} C(t)}{\partial (J_{\beta}^{m,n})^{p-1}}\propto t\frac{\partial^{p-1} E_0}{\partial (J^{m,n}_{\beta})^{p-1}},\\
&\quad\quad\quad\quad\quad\quad\vdots\\
\text{and}\quad&\frac{\partial^{p-1} C(t)}{\partial (J_{\beta}^{m,n})^{p-1}}\propto E_0t,
\end{split}
\end{align}
the $(p-1)$th derivative of the STC with respect to some Hamiltonian parameter is a function of all $q$th derivatives of the ground state energy with respect to the same parameter, for $0\leq q\leq p$. Thus a $p$th order QPT, which corresponds to a discontinuity or divergence of the $p$th derivative of the ground-state energy, can be identified by the $(p-1)$th derivative of the STC.\\

The validity of this result is not affected by taking the expansion of the exponential of Eq.~\eqref{correlation} to higher orders. Consider, for instance, the second-order correction $C^{(2)}(t)$, which adds the terms
\begin{align} \label{second_order}
\begin{split}
C^{(2)}(t)=&-\frac{t^2}{2}\cos(E_0t)\times\\
&\left(E_0^2-2E_0\braGS f_{k}^{\mu}\ketGS + \braGS (f_{k}^{\mu})^2\ketGS\right)
\end{split}
\end{align}
to the time correlations in Eq.~\eqref{time_correl_final}. The term $\braGS f_{k}^{\mu}\ketGS$ has the form already displayed in Eq.~\eqref{time_correl_final}. The third component $\braGS (f_{k}^{\mu})^2\ketGS$ results in more complicated (up to three-site) expectation values in addition to more terms $\partial E_0/\partial J_{\alpha}^{j,k}$ and $\partial E_0/\partial B_{\alpha}^k$. These elements will continue appearing in higher-order expansions, either separately or in expectation values $\braGS f_{k}^{\mu}\ketGS$. So these expansions would lead to the observation of finite-order QPTs as previously discussed for the first-order case.

\bibliography{mybib}	

\end{document}